\documentclass[twocolumn,english,aps,prl,showpacs,superscriptaddress]{revtex4-1}
\usepackage[T1]{fontenc}
\usepackage{amsmath}
\usepackage{amssymb}
\usepackage{graphicx}

\makeatletter
\@ifundefined{textcolor}{}
{%
 \definecolor{BLACK}{gray}{0}
 \definecolor{WHITE}{gray}{1}
 \definecolor{RED}{rgb}{1,0,0}
 \definecolor{GREEN}{rgb}{0,1,0}
 \definecolor{BLUE}{rgb}{0,0,1}
 \definecolor{CYAN}{cmyk}{1,0,0,0}
 \definecolor{MAGENTA}{cmyk}{0,1,0,0}
 \definecolor{YELLOW}{cmyk}{0,0,1,0}
}

\usepackage{amsfonts}%\usepackage[T1]{fontenc}
\@ifundefined{definecolor}
 {\usepackage{color}}{}

\def\be{\begin{equation}}
\def\ee{\end{equation}}
\def\bea{\begin{eqnarray}}          
\def\eea{\end{eqnarray}}
\def\bi{\begin{itemize}}
\def\ei{\end{itemize}}

\makeatother

\usepackage{babel}
\begin{document}

\title{ Topological Order in an Entangled SU(2)$\otimes$XY Spin-Orbital Ring}

\author{Wojciech Brzezicki}
\author{Jacek Dziarmaga}
\affiliation{Marian Smoluchowski Institute of Physics, Jagellonian University,
             Reymonta 4, PL-30059 Krak\'ow, Poland}

\author{Andrzej M. Ole\'{s} }
\affiliation{Marian Smoluchowski Institute of Physics, Jagellonian University,
             Reymonta 4, PL-30059 Krak\'ow, Poland}
\affiliation{Max-Planck-Institut f\"ur Festk\"orperforschung, 
             Heisenbergstrasse 1, D-70569 Stuttgart, Germany}

\date{\today}

\begin{abstract}
We present rigorous topological order which emerges in a 
one-dimensional spin-orbital model due to the ring topology.
Although an exact solution of a spin-orbital ring with SU(2) spin and 
XY orbital interactions separates spins from orbitals by means of a 
unitary transformation, the spins are not independent when the ring 
is closed, but form two half-rings carrying opposite pseudomomenta. 
We show that an inverse transformation back to the physical degrees 
of freedom entangles the spin half-rings with the orbitals once 
again. This surprising correlation arises on changing the topology 
from an open to a closed chain, which reduces the degeneracy of the
ground-state manifold, leaving in it only the states in which
pseudomomenta compensate each other.
\end{abstract}


\maketitle

Spin-orbital physics is one of the foundation stones in the theory 
of frustrated magnetism \cite{Tok00,Hfm,Kha05,Ole05,Ole12}.
When degenerate $3d$ orbitals in a transition-metal oxide are partly
filled, electrons localize due to large on-site Coulomb interaction 
and superexchange between magnetic ions includes both spin and orbital 
degrees of freedom that are strongly interrelated \cite{Kug82}. 
The orbital degeneracy leads in many cases to a dramatic
increase of quantum fluctuation \cite{Fei97}, which may trigger exotic 
order \cite{Brz12} or may stabilize a spin-liquid \cite{Wan09,Nor08} 
when different states compete near a quantum critical point. While 
spin-orbital separation is possible in one-dimensional (1D) systems 
\cite{Woh11}, as observed recently in Sr$_2$CuO$_3$ \cite{Sch12}, 
spins and orbitals are usually entangled strongly, as in the archetypal
Kugel-Khomskii model \cite{Lun12}. In the $S=1/2$ SU(2)$\otimes$SU(2) 
chain \cite{Aff00,vdB98}, both ground state \cite{Che07} and excited 
states \cite{You12} are entangled, similar to the $S=1$ 
SU(2)$\otimes$SU(2) chain which plays a prominent 
role in perovskite vanadates \cite{Kha05,Ole05,Ole12,Kha01}. 
Only in exceptional cases can such 1D models be solved exactly, for 
example at the SU(4) point \cite{su4} or for a valence-bond state 
\cite{Ole07} of alternating spin and orbital singlets \cite{Maj69}, but 
even in these situations the spins and orbitals cannot be separated 
from each other. 

In real materials the symmetry between spin and orbital interactions
is absent. Orbital interactions generically have lower symmetry than 
spin ones \cite{vdB99}, being usually Ising- or XY-like \cite{Pat00}. 
The XY case is quantum and in general the orbitals cannot be separated 
from the spins \cite{Her11}. In this context the 1D spin-orbital 
SU(2)$\otimes$XY model introduced by Kumar \cite{Kum13} is 
surprising --- by a change of basis, the $S=1/2$ spins decouple from 
the orbitals in an {\it open} chain. The orbital interactions remain 
formally unchanged but the spin ones are gauged away. The spins then 
appear free and the ground state has large degeneracy ($2^L$ for chain 
length $L$) \cite{Kum13}.

Frustrated spin systems are at the forefront of modern condensed matter 
theory and experiment \cite{Qsl,Bal10,Nor09}, in large part for the 
investigation of topology in many-body physics. A particular 
manifestation is the topological spin liquid (TSL) \cite{Wen02}, 
a category including resonating valence-bond (RVB) states 
\cite{WhiteKagome} and states hosting excitations with non-Abelian 
fractional quantum statistics, long sought in quantum information and 
topological quantum computation \cite{topoc,Dou05,Tro10} 
(topological protection against decoherence). One experimental example 
of topological order is the fractional quantum Hall effect, where the 
excitations are usually Abelian \cite{Ber06}. The search for realistic 
TSL models gained momentum after the demonstration \cite{WhiteKagome} 
of a Z$_2$ TSL ground state for the $S=1/2$ kagome antiferromagnet. 
Recent progress has been due largely to advanced numerical methods, 
including extension of the quasi-1D density matrix renormalization 
group (DMRG) approach \cite{WhiteKagome} to 2D TSLs \cite{CincioVidal} 
and the development of intrinsically high-dimensional techniques such 
as the projected entangled-pair states (PEPS) ansatz \cite{PEPS} and 
its extension to ``simplex'' lattice units \cite{PESS}. 
PEPS have been used in TSLs for: 
(i) very efficient representation of the RVB state \cite{Poi12}, 
(ii) classifying the topologically distinct ground states of the 
kagome antiferromagnet \cite{PepsKagome}, and 
(iii) demonstrating a TSL in the antiferromagnetic $J_1$-$J_2$ model 
on the square lattice \cite{Wan13}. 
Despite these recent breakthroughs, however, the fingerprints of 
topological order remain notoriously difficult to detect definitively, 
and any exactly solvable model with a TSL ground state \cite{Kit06} 
would be of enormous value.
 
In this Letter we solve exactly the Kumar model \cite{Kum13} on a 
{\it ring} and investigate: 
(i) spin-orbital entanglement, 
(ii) ground state degeneracy, 
(iii) nature of excitations, and 
(iv) scaling of the excitation gap with increasing system size.
We show that, surprisingly, the properties of the Kumar model are 
determined by topology when the closing bond removes the total disorder 
in spins and generates spin-orbital entanglement in the ground state. 

We consider the following model on a {\it ring} of length $L$ with 
periodic boundary condition, 
\begin{equation}
{\cal H}=
\sum_{l=1}^{L}X_{l,l+1}\left(\tau_{l+1}^{+}\tau_{l}^{-}
+\tau_{l+1}^{-}\tau_{l}^{+}\right),
\label{eq:H}
\end{equation}
where 
$X_{l,l+1}\!=(1+\vec{\sigma}_l\cdot\vec{\sigma}_{l+1})/2$ is a 
{\it spin transposition} operator on the bond $\langle l,l+1\rangle$, 
i.e., $X_{l,l+1}\vec{\sigma}_lX_{l,l+1}\!=\!\vec{\sigma}_{l+1}$,
where $\sigma_l$'s are spin Pauli matrices, and $\tau_l$'s are orbital 
Pauli matrices. For an open chain \cite{Kum13}, the spins and orbitals 
are decoupled by a unitary transformation, 
\begin{equation}
{\cal U}=
\prod_{l=1}^{L-1}\left[\frac{1-\tau_{l+1}^{z}}{2}
+\frac{1+\tau_{l+1}^{z}}{2}\chi_{l+1,l}\right],
\label{eq:U}
\end{equation}
where $\chi_{l+1,l}$ is a {\it spin permutation} operator composed of 
the spin transpositions $X_{i,j}$: 
\begin{equation}
\chi_{l+1,l}= X_{l+1,l}X_{l,l-1}...X_{3,2}X_{2,1}.
\end{equation}
For a periodic chain, the same ${\cal U}$ maps the model (\ref{eq:H}) to 
\begin{equation}
\tilde{{\cal H}}=
{\cal U}^{\dagger}{\cal H}\:{\cal U} = 
\left(
\sum_{l=1}^{L-1}\tau_{l+1}^{+}\tau_{l}^{-} 
+ R_{1}^{(1)}R_{1}^{(2)}\tau_{1}^{+}\tau_{L}^{-}
\right)+
{\rm h.c.}.
\label{eq:HRR}
\end{equation}
Here $R_{p}^{(1)}$ is a cyclic permutation of spins at sites 
$l=1,\dots,N$ by $p$ sites, 
$
R_{p}^{(1)} \vec{\sigma_l} R_{p}^{(1)\dag} = \vec{\sigma}_{l+p}
$,
and $R_{p}^{(2)}$ is the same permutation at sites $l=(N\!+\!1),...,L$.
Thereby the ring separates into two parts of length $N$ and $(L-N)$, 
with $N$ being a good quantum number of ``up'' orbitals,
\begin{equation}
N=\frac12\sum_{l=1}^{L}(1+\tau_{l}^{z}).
\label{N}
\end{equation} 
For more details on the derivation of Eq. (\ref{eq:HRR}) see
the Supplemental Material \cite{suppl}.

\begin{figure}[t!]
\begin{centering}
\includegraphics[clip,width=1\columnwidth]{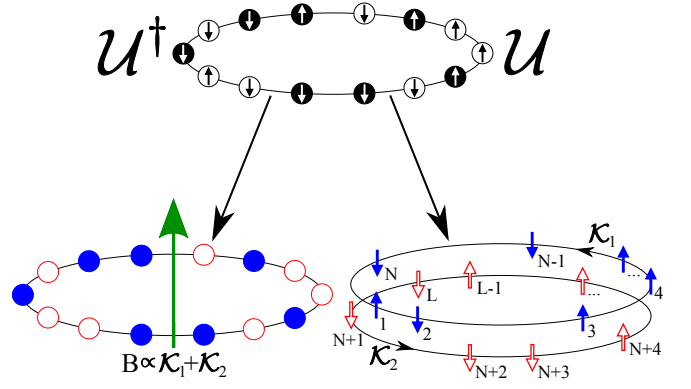} 
\par\end{centering}
\caption{(color online) 
Artist's view of the spin-orbital decoupling in the ring Eq. 
(\ref{eq:H}) caused by the transformation ${\cal U}$. 
The initial spin-orbital chain (top) splits into purely orbital (left) 
and spin (right) segments. The orbital part feels an external magnetic 
field $\vec{B}$ perpendicular to the ring (arrow) produced by the spin 
flows ${\cal K}_1$ and ${\cal K}_2$.
}
\label{fig:kum_dec} 
\end{figure}

Unlike in an open chain, the spins are not fully integrated out but 
show up at the closing bond $\langle L,1\rangle$ of the Hamiltonian 
(\ref{eq:HRR}). The unitary operator $R_{1}^{(1)}$ has 
eigenvalues $e^{i{\cal K}_{1}}$ with a quasimomentum 
${\cal K}_{1}\!=\!2\pi n_{1}/N$ and $n_{1}=0,...,N\!-\!1$. Similarly, 
for $R_{1}^{(2)}$ we get ${\cal K}_{2}\!=\!2\pi n_{2}/(L\!-\!N)$ and 
$n_{2}\!=\!0,...,L\!-\!N\!-\!1$. Thus the spin sector enters 
$\tilde{{\cal H}}$ as a single phase factor on the $\langle L,1\rangle$ 
bond,
\begin{equation}
\tilde{{\cal H}}=
\sum_{l=1}^{L-1}\left(\tau_{l+1}^{+}\tau_{l}^{-}
 +  e^{i\left({\cal K}_{1}+{\cal K}_{2}\right)}\tau_{L}^{+}\tau_{1}^{-}
 +{\rm h.c.}\right).
\label{eq:HRRbis}
\end{equation}
It can be diagonalized by a Jordan-Wigner (JW) transformation,
$
\tau_{l}^{z}=1-2n_{l}, 
\tau_{l}^{+}=c_{l}\prod_{j<l}\,(1-2n_{j}),
$
where $c_{l}$ annihilates a JW fermion and 
$n_{l}=c_{l}^{\dagger}c_{l}^{}$:  
\begin{equation}
\tilde{{\cal H}}=\sum_{l=1}^{L-1}\left(c_{l+1}^{\dagger}c_{l}^{}
+e^{-2\pi i\Phi}c_{1}^{\dagger}c_{L}^{}+{\rm h.c.}\right)=
2\sum_{k}c_{k}^{\dagger}c^{}_{k}\cos k.
\label{eq:tildeH}
\end{equation}
Here the phase $\Phi=n_{1}/N+n_{2}/(L\!-\!N)-(L\!-\!N\!-\!1)/2$ 
is twisting the boundary condition, $c_{L+1}=e^{2\pi i\Phi}c_{1}$, 
just like an Aharonov-Bohm magnetic flux $\Phi$ through the periodic 
ring. The schematic view of the spin-orbital decoupling is shown in 
Fig. \ref{fig:kum_dec}. Note that correlated states are also found in 
the 1D XY$\otimes$XY spin-orbital model \cite{Mil99}, but here their 
properties are more subtle, see below.

%%%%%%%%%%%%%%%%%%%%%%%%%%%%%%%%%%%%%%%%%%%%%%%%%%%%%%%%%%%%%%%%%%%%%%%
%%%%%%%%%%%%%%%%%%%%%%%%%%%%%%%%%%%%%%%%%%%%%%%%%%%%%%%%%%%%%%%%%%%%%%%

The quasimomenta in Eq. (\ref{eq:tildeH}) are quantized as 
$k=2\pi\left(m\!+\!\Phi\right)/L$, where $m=0,...,(L\!-\!1)$. Assuming 
that $L$ is even, $\tilde{{\cal H}}$ is minimized by $N=L/2$ and
\begin{equation}
{\cal K}\equiv{\cal K}_{1}+{\cal K}_{2}=0,
\end{equation}
meaning that half of the orbitals are up, or the Fermi sea $|O\rangle$ 
is half-filled with the JW fermions. At the same time, there are two 
anticorrelated flows in the spin half-chains: when the first one has a 
quasimomentum ${\cal K}_{1}$, 
the second one has the opposite quasimomentum ${\cal K}_2=-{\cal K}_1$.

\begin{figure}[t!]
\begin{centering}
\includegraphics[clip,width=1\columnwidth]{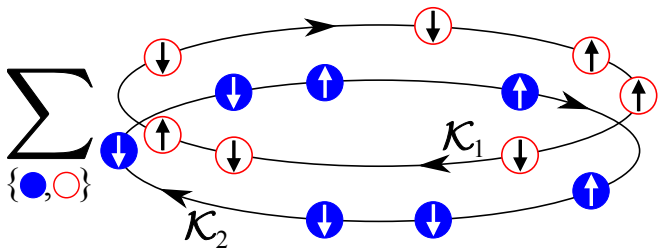} 
\par\end{centering}
\caption{(color online) 
Schematic view of the ground state of the model (\ref{eq:H}). The
system is divided into orbital "up" (empty circles) and orbital "down" 
(full circles) subsystems carrying the spin flows ${\cal K}_1$ and 
${\cal K}_2$. The flows are synchronized in this representation
in each component: ${\cal K}_1={\cal K}_2$. Despite this nonlocal 
correlation, the local orientation of individual spins is (almost) 
random. Here the sum runs over all orbital configurations with the 
same number $N=L/2$ of "up" and "down" orbitals with amplitudes that 
are omitted here.
}
\label{fig:kum_so} 
\end{figure}

To see the action of the transformation back to the original basis more 
clearly, we represent the orbital Fermi sea in the $\tau^z$-eigenbasis 
as $|O\rangle=
\sum_{\vec{\alpha}}O_{\vec{\alpha}}\left|\vec{\alpha}\right\rangle$, 
where $\tau_{l}^{z}\left|\vec{\alpha}\right\rangle= 
\alpha_{l}\left|\vec{\alpha}\right\rangle$,
which gives in the original basis, 
\begin{equation}
{\cal U}|O\rangle=\sum_{\vec{\alpha}}O_{\vec{\alpha}} 
\left|\vec{\alpha}\right\rangle {\cal U}_{\vec{\alpha}}.
\label{Ualpha}
\end{equation}
The orbital state $\left|\vec{\alpha}\right\rangle $ is not altered, 
but the spins are subject to a permutation ${\cal U}_{\vec{\alpha}}$ 
that maps the spins $\{N,...,1\}$ to the successive $N$ sites with 
orbitals "up" ($\alpha_l=1$), and the spins $\{N\!+\!1,...,2N\}$ to the 
remaining $N$ successive sites with orbitals "down" ($\alpha_l=-1$). 
As the sequence of spins $\{1,...,N\}$ is {\it reversed} by the 
transformation, the spin-flow anticorrelation transforms into a 
correlation between the spin flow ${\cal K}_1$ on sites with orbitals 
up and the flow ${\cal K}_2={\cal K}_1$ on sites with orbitals down. 
The spin-orbital entanglement in the ground state wave functions can 
be depicted as in Fig. \ref{fig:kum_so}, where first the purely 
orbital Fermi sea wave function $|O\rangle$ is decomposed in the 
$\tau_i^z$ basis, and then the "up" and "down" components are dressed 
with equal spin flows $\{{\cal K}_1,{\cal K}_2\}$. This demonstrates
the spin-orbital entanglement present in all the quasimomentum states 
which form the ground state.

%%%%%%%%%%%%%%%%%%%%%%%%%%%%%%%%%%%%%%%%%%%%%%%%%%%%%%%%%%%%%%%%%%%%%%%
%%%%%%%%%%%%%%%%%%%%%%%%%%%%%%%%%%%%%%%%%%%%%%%%%%%%%%%%%%%%%%%%%%%%%%%

\begin{figure}[t!]
\begin{centering}
\includegraphics[clip,width=5.2cm]{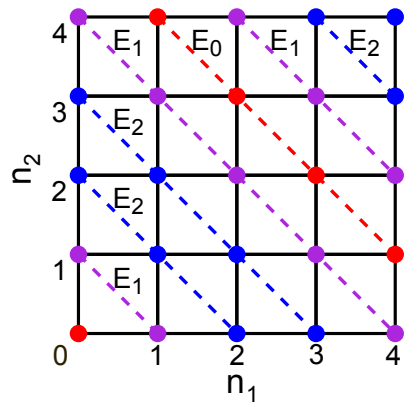} 
\par\end{centering}
\caption{(color online) 
Energies of the lowest topological excitations for $L=10$ and $N=L/2=5$ 
in the periodic $(n_1,n_2)$ reciprocal plane. Dots connected by dashed 
lines represent different energies $E_m$ --- the lines of constant 
$n=0,1,2$, where $n=n_{1}+n_{2}$, are shown as a guide for the eye.
}
\label{fig:kum_topo}
\end{figure}

Among the excited states one has to distinguish between the orbital 
excited states of the chain (\ref{eq:tildeH}) for a fixed ${\cal K}$, 
and the excitations with energy increased because of finite ${\cal K}$, 
or asynchronized flows ${\cal K}_1,{\cal K}_2$ in Fig. \ref{fig:kum_so}. 
The latter are of interest here as they arise from the degenerate 
manifold of $2^L$ ground states of an open chain and gain finite 
dispersion only due to the change of topology caused by the closing bond 
--- they can be regarded as {\it topological excitations}. The energy of 
such excitations depends on total quasimomentum ${\cal K}=2\pi n/N\neq0$ 
that enters the orbital Hamiltonian (\ref{eq:tildeH}). Thus
every ${\cal K}$-excited state is related with a different orbital 
ground state where all the quasimomenta $k$ from the Fermi sea gain 
a shift of $\delta k={\cal K}/L$ with respect to the global ground
state. This shows both non-local and entangled nature of the
topological excitations which change the global spin flows and shift 
the orbital Fermi sea at the same time. The multiplet structure of the 
${\cal K}$-excited states is such that the first excited state has 
$n=\pm 1$ corresponding to ${\cal K}=\pm 2\pi/N$. The second excited 
state is obtained with $n=\pm2$ and so on, see Fig. \ref{fig:kum_topo} 
for classification of the excitations. 
Their dispersion is quadratic in $n$ for $|n|\ll L/2$ and large $L$,
\begin{equation}
{\cal E}_n=-\frac{2L}{\pi}+16\pi\frac{n^2}{L^3}.
\end{equation}
The energy gap between the orbital ground states for $n=0$ and $n=\pm1$ 
for nanoscopic systems. It scales as $L^{-3}$, while the orbital gap 
gradually closes with increasing system size, and
for a given $n$ decays as $L^{-1}$.

We now solve the problem of the ground state degeneracy for the 
spin-orbital SU(2)$\otimes$XY ring Eq. (\ref{eq:H}). An eigenstate 
of $R_{1}^{(1)}$ with eigenvalue $e^{i{\cal K}_1}$ can be generated 
from the $m_1$-th basis state $|m_1\rangle$ of spins $1,..,N$ as
\begin{equation}
\left|{\cal K}_1,m_1\right\rangle =\frac{1}{\sqrt{N}}
\sum_{p=0}^{N-1}e^{-i{\cal K}_1 p} R_{p}^{(1)}
|m_1\rangle.
\label{eq:stK1}
\end{equation}
Analogically, one gets the $e^{i{\cal K}_{2}}$-eigenstate for spins 
labeled as $(N+1),..,2N$. Note that the basis states $|m_i\rangle$ that 
are periodic under cyclic permutations $R^{(i)}_{p}$ with $p<N$ can 
generate ${\cal K}_i=2\pi n/p$, with $n=0,...,(p-1)$ only. For instance, 
the ferromagnetic (FM) state can generate only ${\cal K}_i\!=\!0$ and 
the anti-FM one ${\cal K}_i\!=\!0,\pi$. In case of a prime $N$, the 
only periodic states are the two FM states. Thus there are 
${\cal N}\!=\!(2^{N}\!-\!2)/N$ quasimomentum-degenerate eigenstates 
when ${\cal K}_i\!\neq\!0$, and their number is ${\cal N}\!+\!2$ for 
${\cal K}_i\!=\!0$. The total ground state degeneracy is 
\begin{equation}
{\cal D}=(N-1){\rm {\cal N}}^2+({\cal N}+2)^{2}.
\end{equation}
For a large system size one finds ${\cal D}\approx2^{L+1}/L$ as 
compared to ${\cal D}_0=2^L$ for an open chain. Thus, the degeneracy 
is drastically reduced by the topological correlations introduced 
by the periodic boundary conditions.

%%%%%%%%%%%%%%%%%%%%%%%%%%%%%%%%%%%%%%%%%%%%%%%%%%%%%%%%%%%%%%%%%%%%%%%
%%%%%%%%%%%%%%%%%%%%%%%%%%%%%%%%%%%%%%%%%%%%%%%%%%%%%%%%%%%%%%%%%%%%%%%%

To understand the structure of the degenerate ground state, we analyze 
now the spin and orbital part. 
Due to the degeneracy, the zero temperature state is a mixed state. 
In the new basis, its density matrix is a product 
$\tilde{\rho}=\left|O\right\rangle\left\langle O\right|\rho_{S}$, where 
the spin state is 
\begin{equation}
\rho_S=\frac{1}{\cal D}\sum_{{\cal K}}
\sum_{m_{1}m_{2}}\!\left|-{\cal K},m_{1}
\right\rangle \left\langle -{\cal K},m_{1}\right|~
\left|{\cal K},m_{2}\right\rangle \left\langle {\cal K},m_{2}\right|.
\label{rhoS}
\end{equation}
Back in the physical representation, the product state 
$\rho={\cal U}\,\tilde{\rho}\,{\cal U}^{\dagger}$ becomes spin-orbital 
entangled by the inverse transformation (\ref{Ualpha}). In the orbital 
sector, this transformation is a pure decoherence in the pointer basis 
of $\tau_{l}^{z}$ \cite{Zur80}. In this basis the orbital state 
$|\vec\alpha\rangle$ does not change, but the spins are subject to a 
transformation ${\cal U}_{\vec\alpha}$ that depends on the orbital state:
\begin{eqnarray}
\rho &=& \frac{1}{\cal D}\,\sum_{\vec{\alpha},\vec{\beta}}\,
O_{\vec{\alpha}}^{}O_{\vec{\beta}}^{*}~\sum_{{\cal K},m_{1},m_{2}}  
 \left|\vec{\alpha}\right\rangle~ 
 {\cal U}_{\vec{\alpha}}\left|-{\cal K},m_{1}\right\rangle 
 \left|{\cal K},m_{2}\right\rangle \nonumber \\
& \times & \left\langle-{\cal K},m_{1}\right|
 \left\langle {\cal K},m_{2}\right|
 {\cal U}_{\vec{\beta}}^{\dagger}~\langle\vec{\beta}|~. 
\end{eqnarray}
%%AM
Thus
the orbital state $\left|\vec{\alpha}\right\rangle $ becomes 
entangled with spin states ${\cal U}_{\vec{\alpha}}
\left|-{\cal K},m_{1}\right\rangle\left|{\cal K},m_{2}\right\rangle$, 
and with a flow ${\cal K}$ 
at sites $\alpha_l=1$ and another flow ${\cal K}$ at sites 
$\alpha_l=-1$. The entanglement has decoherence effect, 
\begin{equation}
\rho_{O}={\rm Tr}_{S}\rho=\sum_{\vec{\alpha},\vec{\beta}}
\left|\vec{\alpha}\right\rangle ~O_{\vec{\alpha}}^{}D_{\vec{\alpha},\vec{\beta}}O_{\vec{\beta}}^{*}~\langle\vec{\beta}|~,
\end{equation}
comparing with the pure orbital state before the inverse 
${\cal U}$-transformation,
$\left|O\right\rangle \left\langle O\right|=
\sum_{\vec{\alpha},\vec{\beta}}\left|\vec{\alpha}\right\rangle ~O_{\vec{\alpha}}^{}O_{\vec{\beta}}^{*}~\langle\vec{\beta}|$. 
The coherences are suppressed by the decoherence factors,
\begin{equation}
D_{\vec{\alpha},\vec{\beta}}={\rm Tr}_{S}
~{\cal U}_{\vec{\alpha}}\rho_{S}\:{\cal U}_{\vec{\beta}}^{\dagger},
\label{eq:D}
\end{equation}
but the probabilities $O_{\vec{\alpha}}^{}O_{\vec{\alpha}}^{*}$ are 
invariant, $D_{\vec{\alpha},\vec{\alpha}}=1$. Since $\rho_S$ in 
(\ref{rhoS}) is a diagonal ensemble of spin ground states 
$|S\rangle\langle S|$, $D_{\vec{\alpha},\vec{\beta}}$ measures how 
much orthogonal the states ${\cal U}_{\vec\alpha}|S\rangle$ and 
${\cal U}_{\vec\beta}|S\rangle$ are or, equivalently, how much 
entangled with spins the orbital states 
$|\vec\alpha\rangle$ and $|\vec\beta\rangle$ become for a typical 
$|S\rangle$. 

Since the $\tau^z$-basis is the basis of JW fermionic occupation numbers, 
the decoherence is localizing the JW fermions that were originally 
delocalized in the Fermi sea $|O\rangle$. To get an idea how tight the 
localization is, we can consider an open chain where the closing bond 
$\langle L,1\rangle$ is missing and, consequently, the spin state is 
arbitrary, $\rho_{S}\propto 1$. One finds a compact formula, 
\begin{equation}
\log_{2}D_{\vec{\alpha},\vec{\beta}}=
-L+\sum_{p=1}^{L}\delta_{0,\sum_{l=1}^{p}(\alpha_{l}-\beta_{l})}.
\label{eq:Dopen}
\end{equation}
An illustrative example are states $\vec{\alpha}$ and 
$\vec{\beta}$ which differ at only two sites $\{i,j\}$. Then 
$D_{\vec\alpha,\vec\beta}=2^{-|i-j|}$ is localized exponentially 
on the scale of $(\ln 2)^{-1}=1.44$ sites. 

The short localization length suggests that Eq. (\ref{eq:Dopen}) is 
a good approximation in the periodic case as well. 
Indeed, in case of prime $N$ one finds that 
\begin{equation}
D_{\vec{\alpha},\vec{\beta}}=\frac{1}{N{\cal D}}\sum_{p=0}^{N-1}
{\rm Tr}_{S}~{\cal U}_{\vec{\beta}}^{\dagger}\:{\cal U}_{\vec{\alpha}}
R_{-p}^{(1)}R_{-p}^{(2)}=
\frac{1}{N{\cal D}}\sum_{p=0}^{N-1}2^{c_{p}}.
\label{eq:Calphabeta}
\end{equation}
Here $c_{p}$ is a number of cycles in a cyclic decomposition of the 
permutation 
${\cal U}_{\vec{\beta}}^{\dagger}\:
{\cal U}_{\vec{\alpha}}^{}R_{-p}^{(1)}R_{-p}^{(2)}$.
Again, for $\vec{\alpha}$ and $\vec{\beta}$ that differ only at sites 
$\{i,j\}$ without loss of generality we set $i\!=\!1$, $j\!=\!1\!+\!R$, 
with $1\!\leq\! R\!\leq\! N$, and $\alpha_{1}\!=\!\beta_{1+R}\!=\!1$, 
$\alpha_{1+R}\!=\!\beta_{1}\!=\!-1$,
and introduce integer sums,
\begin{equation}
a_{\pm}=\frac12\sum_{l=2}^{R}(1\pm\alpha_l),\hskip .5cm
b_{\pm}=\frac12\sum_{l=R+2}^{2N}(1\pm\alpha_l).
\end{equation}
They can be decomposed as $a_{\pm}\!=\!C_{\pm}(p-1)\!
+\!A_{\pm}$ and $b_{\pm}\!=\!(p-1)\!+\!D_{\pm}p+\!B_{\pm}$,
where $C_{\pm},D_{\pm}$ are integers and $A_{\pm},B_{\pm}$ 
are integer remainders: $0\leq A_{\pm}<(p-1)$ and $0\leq B_{\pm}<p$. 
The numbers of cycles $c_p$ can now be written in a more compact form. 
In particular, $c_{0}\!=\!2N\!-\!R$, $c_{1}\!=\!R$, 
and for $p\geq2$ we obtain
\begin{equation}
c_{p\geq2}\!=\!{\cal C}\left[R_{p,B_{+}}R_{p-1,A_{+}}\right]
\!+\!{\cal C}\left[R_{p,B_{-}}R_{p-1,A_{-}}\right] \!-\!1\,.
\label{cp2}
\end{equation}
Here $R_{p,B_{+}}$ is a cyclic permutation of a $p$-element list by 
$B_{+}$ sites, and $R_{p-1,A_{+}}$ refers to the first $(p\!-\!1)$ 
elements of this list. The function ${\cal C}$ counts the number of 
cycles. From Eq. (\ref{cp2}) we can estimate that 
%%AM
%$c_{p\geq2}\!\leq\!2p\!-\!1$. Since $c_{N-p}\!=\!c_{p}$ we also have 
$c_{p\geq2}\!\leq\!2p\!-\!1$. Since $c_{N-p}\!=\!c_{p}$, it follows 
$c_{p\geq2}\!\leq\! 2(N\!-\!1)/2\!-\!1\!=\!N\!-\!2$. For large $N$ 
the sum (\ref{eq:Calphabeta}) is dominated by the $p=0$ term,
\begin{equation}
D_{\vec{\alpha},\vec{\beta}}\approx
\frac{1}{N{\cal D}}\!\left[2^{2N-R}\!+\!2\times2^{R}
\!+\!{\cal O}\left(N2^{N}\right)\right]\approx2^{-R}\,,
\end{equation}
with the same localization length as in the open chain. This result 
illustrates the general smallness of decoherence factors corroborating 
the picture of non-trivial spin-orbital entanglement represented in 
Fig. \ref{fig:kum_so}. For more details see the Supplemental Material
\cite{suppl}.

Summarizing, we have shown rigorously that closing the 
spin-orbital chain with SU(2)$\otimes$XY exchange Eq. (\ref{eq:H})
causes surprising changes in the spin part of the lowest-lying 
eigenstates. Spins are not decoupled from the orbitals, as it happens 
in the open chain, but instead the spin states associated with the 
orbital ground state are structured in a multiplet labeled by their 
quasimomentum. We have found that this change of the exact eigenstates:  
(i) reduces the degeneracy of the ground state and, more importantly, 
(ii) triggers nontrivial topological order with spin-orbital 
entanglement. 

Similar entanglement concerns also the excited states ---
spin excitations have definite quasimomenta on sites where the 
orbitals are polarized up and on those polarized down, and the total 
quasimomentum is a good topological quantum number. These topological 
excitations have a dispersion quadratic in the quasimomentum and a 
nontrivial gap scaling exponent: $\Delta\propto L^{-\eta}$ with 
$\eta= 3$. The orbital decoherence caused by the inverse 
${\cal U}$-transformation, that entangles spins with orbitals, 
localizes the orbital quasiparticles on a very short length scale, 
both for periodic and open chains.

%%%%%%%%%%%%%%%%%%%%%%%%%%%%%%%%%%%%%%%%%%%%%%%%%%%%%%%%%%%%%%%%%%%%%%%

We thank particularly warmly Bruce Normand for valuable advice, and 
Giniyat Khaliullin and Krzysztof Wohlfeld for insightful discussions. 
We acknowledge financial support by the Polish National Science Center 
(NCN) under Projects No. 
2012/04/A/ST3/00331 (W.B. and A.M.O) and 
2011/01/B/ST3/00512 (J.D.).

%%%%%%%%%%%%%%%%%%%%%%%%%%%%%%%%%%%%%%%%%%%%%%%%%%%%%%%%%%%%%%%%%%%%%%%
\vfill

\eject

\section*{Supplemental Material}

In the first Section of this Supplemental Material we present the 
derivation of the unitary transformation on the closing bond in the 
spin-orbital model introduced by Kumar 
[B. Kumar, Phys. Rev. B \textbf{87}, 195105 (2013)], 
see Eq. (1) of the Letter. 
In the second Section we show the technical details which justify
our analysis of the structure of the correlated states which form 
the degenerate ground state of the closed SU(2)$\otimes$XY 
spin-orbital ring.

%%%%%%%%%%%%%%%%%%%%%%%%%%%%%%%%%%%%%%%%%%%%%%%%%%%%%%%%%%%%%%%%%%%%%%%
\subsection{The closing bond in the SU(2)$\otimes$XY Hamiltonian}
%%%%%%%%%%%%%%%%%%%%%%%%%%%%%%%%%%%%%%%%%%%%%%%%%%%%%%%%%%%%%%%%%%%%%%%

Here we derive the result of the unitary transformation
\begin{eqnarray}
{\cal U} &=&
\prod_{l=1}^{L-1}
\left[
\frac{1-\tau_{l+1}^{z}}{2}+
\frac{1+\tau_{l+1}^{z}}{2}\chi_{l+1,l}
\right]\nonumber\\
&\equiv&
\prod_{l=1}^{L-1}
\left[
P^-_{l+1}+P^+_{l+1}\chi_{l+1,l}
\right],
\label{eq:U:app}
\end{eqnarray}
on the periodic Kumar Hamiltonian
\begin{equation}
{\cal H}=\sum_{l=1}^{L}X_{l,l+1}\left(\tau_{l+1}^{+}\tau_{l}^{-}
+\tau_{l+1}^{-}\tau_{l}^{+}\right).
\label{eq:H:app}
\end{equation}
We focus on the closing bond $\langle L,1 \rangle$ in Eq. (2). 
It is enough to consider the transformation,
\begin{eqnarray}
&& {\cal U}^\dag~X_{L,1}\tau_{1}^{+}\tau_{L}^{-}~{\cal U} = \nonumber\\
&& 
\prod_{l=L-1}^{1}
\left[
P^-_{l+1}+P^+_{l+1}\chi_{l+1,l}^\dag
\right]\times
\nonumber\\
&&
X_{L,1}\tau_{1}^{+}\tau_{L}^{-}~
\prod_{l'=1}^{L-1}
\left[
P^-_{l'+1}+P^+_{l'+1}\chi_{l'+1,l'}
\right] =
\nonumber\\
&&
\prod_{l=L-1}^{1}
\left[
P^-_{l+1}+P^+_{l+1}\chi_{l+1,l}^\dag
\right]\times 
\nonumber\\
&&
X_{L,P^+_1}~\tau_{1}^{+}\tau_{L}^{-}
\prod_{l'=1}^{L-1}
\left[
P^-_{l'+1}+P^+_{l'+1}\chi_{l'+1,l'}
\right]=
\nonumber\\
&&
\prod_{l=L-1}^{2}
\left[
P^-_{l+1}+P^+_{l+1}\chi_{l+1,l}^\dag
\right]\times 
\nonumber\\
&&
X_{L,P^+_1+P^+_2}~\tau_{1}^{+}\tau_{L}^{-}
\prod_{l'=2}^{L-1}
\left[
P^-_{l'+1}+P^+_{l'+1}\chi_{l'+1,l'}
\right]=
\nonumber\\
&&
\left[
P^-_{L}+P^+_{L}\chi_{L,L-1}^\dag
\right]\times 
\nonumber\\
&&
X_{L,\sum_{l=1}^{L-1}P^+_l}~\tau_{1}^{+}\tau_{L}^{-}
\left[
P^-_{L}+P^+_{L}\chi_{L,L-1}
\right].
\end{eqnarray}
After further transformations one finds that 
\begin{eqnarray}
&& {\cal U}^\dag~X_{L,1}\tau_{1}^{+}\tau_{L}^{-}~{\cal U} = 
\nonumber\\
%&&
%\prod_{l=L-1}^{1}
%\left[
%P^-_{l+1}+P^+_{l+1}\chi_{l+1,l}^\dag
%\right]\times 
%\nonumber\\
%&&
%X_{L,P^+_1}~\tau_{1}^{+}\tau_{L}^{-}
%\prod_{l'=1}^{L-1}
%\left[
%P^-_{l'+1}+P^+_{l'+1}\chi_{l'+1,l'}
%\right]=
%\nonumber\\
%&&
%\prod_{l=L-1}^{2}
%\left[
%P^-_{l+1}+P^+_{l+1}\chi_{l+1,l}^\dag
%\right]\times 
%\nonumber\\
%&&
%X_{L,P^+_1+P^+_2}~\tau_{1}^{+}\tau_{L}^{-}
%\prod_{l'=2}^{L-1}
%\left[
%P^-_{l'+1}+P^+_{l'+1}\chi_{l'+1,l'}
%\right]=
%\nonumber\\
%&&
%\left[
%P^-_{L}+P^+_{L}\chi_{L,L-1}^\dag
%\right]\times 
%\nonumber\\
%&&
%X_{L,\sum_{l=1}^{L-1}P^+_l}~\tau_{1}^{+}\tau_{L}^{-}
%\left[
%P^-_{L}+P^+_{L}\chi_{L,L-1}
%\right]=
%\nonumber\\
&&
X_{L,\sum_{l=1}^{L-1}P^+_l}~\tau_{1}^{+}\tau_{L}^{-}~
\chi_{L,L-1}=
\nonumber\\
&&
X_{L,\sum_{l=1}^{L}P^+_l}~\tau_{1}^{+}\tau_{L}^{-}~
\chi_{L,L-1}=
\nonumber\\
&&
X_{L,N}\chi_{L,L-1}~\tau_{1}^{+}\tau_{L}^{-}=
\nonumber\\
&&
X_{L,N}X_{L,L-1}...X_{2,1}~\tau_{1}^{+}\tau_{L}^{-}.
\end{eqnarray}
The operator $R=X_{L,N}~X_{L,L-1}...X_{2,1}$ is a spin permutation:
\begin{eqnarray}
&&R\vec{\sigma}_1R^\dag=\vec{\sigma}_N,~~R\vec{\sigma}_2R^\dag=\vec{\sigma}_1,....,
~~R\vec{\sigma}_NR^\dag=\vec{\sigma}_{N-1},~\nonumber\\
&&R\vec{\sigma}_{N+1}R^\dag=\vec{\sigma}_L,~R\vec{\sigma}_{N+2}R^\dag=
\vec{\sigma}_{N+1},....,~R\vec{\sigma}_LR^\dag=\vec{\sigma}_{L-1},~\nonumber
\end{eqnarray}
i.e., it is a cyclic permutation of spins $1,...,N$ by one site, and 
the same permuation of spins $(N+1),...,L$. In other words, the permutation 
$R$ factorizes into two cycles, $R=R_1^{(1)}R_1^{(2)}$, where
\begin{eqnarray}
R_1^{(1)} &=& X_{N,N-1}...X_{2,1}, \nonumber\\
R_1^{(2)} &=& X_{L,L-1}...X_{N+2,N+1}.
\end{eqnarray}
Finally, the closing bond becomes
\begin{equation}
{\cal U}^\dag X_{L,1}\tau_{1}^{+}\tau_{L}^{-} {\cal U}+{\rm h.c.} =
R_1^{(1)}R_1^{(2)} \tau_{1}^{+}\tau_{L}^{-} +{\rm h.c.},
\end{equation}
which justifies Eq. (4) of the Letter.

%%%%%%%%%%%%%%%%%%%%%%%%%%%%%%%%%%%%%%%%%%%%%%%%%%%%%%%%%%%%%%%%%%%%%%%
%%%%%%%%%%%%%%%%%%%%%%%%%%%%%%%%%%%%%%%%%%%%%%%%%%%%%%%%%%%%%%%%%%%%%%%
\subsection{Permutation cycles and decoherence factors}
%%%%%%%%%%%%%%%%%%%%%%%%%%%%%%%%%%%%%%%%%%%%%%%%%%%%%%%%%%%%%%%%%%%%%%%
%%%%%%%%%%%%%%%%%%%%%%%%%%%%%%%%%%%%%%%%%%%%%%%%%%%%%%%%%%%%%%%%%%%%%%%

\begin{figure}[t!]
\begin{centering}
\vspace*{0.5cm}
\includegraphics[clip,width=1\columnwidth]{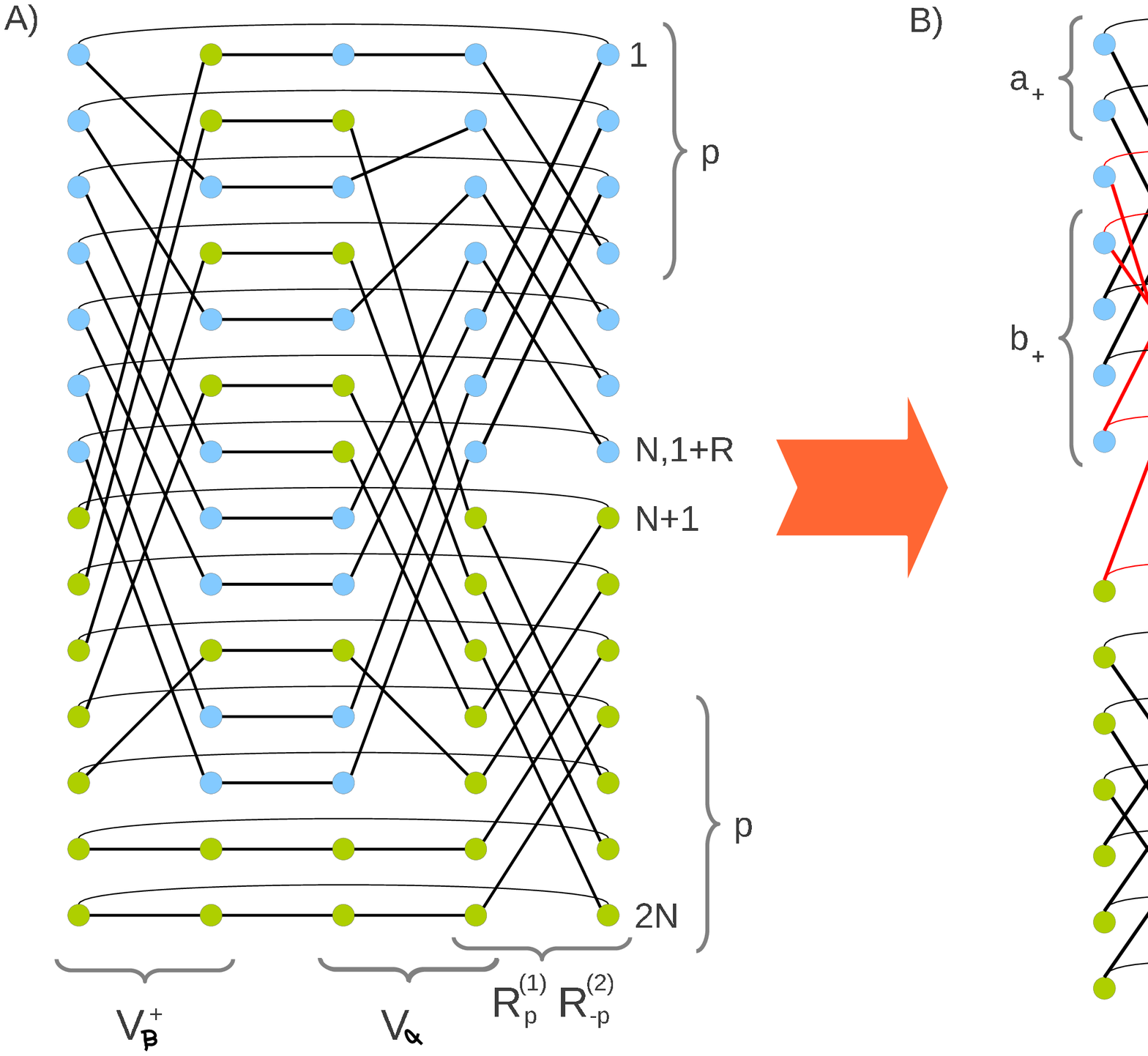} 
\par\end{centering}
\vspace*{-0.2cm}
\caption{
In A, graphic representation of ${\rm Tr}\left[{\cal V}_{\vec{\beta}}^{\dagger}\:{\cal V}_{\vec{\alpha}}R_{p}^{(1)}R_{-p}^{(2)}\right]$.
Here each circle represents a spin. The blue circles represent spins at 
the sites $1,...,N$ that are mapped by ${\cal V}_{\vec\alpha}$ 
(or ${\cal V}_{\vec\beta}$) 
to the sites with orbitals up, where $\alpha_l=1$ (or $\beta_l=1$). 
The green ones represent spins $(N+1),...,2N$ that are mapped to the sites 
with orbitals down. The circles/spins are connected by lines that 
represent actions of the permutations on each spin. The connecting lines 
make $c_p$ closed loops. Each 
loop contributes a factor of $2$ to the trace represented by graph A. 
Here we chose $N=7$, $R=6$, $p=4$, $a_+=2$, $b_+=4$, $a_-=3$, $b_-=3$ 
for illustration, but the argument can be generalized to arbitrary 
values of the parameters. In B, after the three middle spin columns in 
graph A are elliminated, we obtain graph B with the same number of loops.
This graph can be divided into the top half (spins $1,...,N$/blue circles) 
and the bottom one (spins $N+1,...,2N$/green circles).
Notice that there is exactly one closed loop (marked red) that is 
common to the two halves.
In the next Fig. \ref{fig:diagram2} we continue with the top half, 
the analysis of the bottom one follows the same lines, but with 
$\{a_+,b_+\}$ replaced by $\{a_-,b_-\}$.
}
\label{fig:diagram1} 
\end{figure}
%%%%%%%%%%%%%%%%%%%%%%%%%%%%%%%%%%%%%%%%%%%%%%%%%%%%%%%%%%%%%%%%%%%%%%%

We need a more efficient formula for the number of cycles $c_p$ 
in the cycle decomposition of the permutation
$
{\cal U}_{\vec{\beta}}^{\dagger}\:{\cal U}_{\vec{\alpha}}~R_{-p}^{(1)}R_{-p}^{(2)}
$:
\begin{equation}
c_p=
{\cal C}
\left(
{\cal U}_{\vec{\beta}}^{\dagger}\:{\cal U}_{\vec{\alpha}}~R_{-p}^{(1)}R_{-p}^{(2)}
\right).
\end{equation}
Notice that the number of cycles is invariant under cyclic permutations:
\begin{equation}
{\cal C}
\left(
{\cal U}_{\vec{\beta}}^{\dagger}\:{\cal U}_{\vec{\alpha}}~R_{-p}^{(1)}R_{-p}^{(2)}
\right)=
{\cal C}
\left(
{\cal U}_{\vec{\alpha}}~R_{-p}^{(1)}R_{-p}^{(2)}{\cal U}_{\vec{\beta}}^{\dagger}
\right)=....
\end{equation}
just like the trace operation.

The transformation ${\cal U}_{\vec{\alpha}}$ maps the spins $N,...,1$ 
to the spins at the $N$ successive sites with orbitals up, where 
$\alpha_l=1$, and the spins $N+1,...,2N$ to the spins at the remaining 
$N$ successive sites with orbitals down, where $\alpha_l=-1$. 
It is convenient to make a decomposition: 
\begin{equation}
{\cal U}_{\vec{\alpha}}={\cal V}_{\vec{\alpha}}{\cal P}^{(1)}.
\end{equation} 
Here ${\cal P}^{(1)}$ is a parity operation that maps the spins 
$1,...,N$ to the spins $N,...,1$, i.e., inverts the order of spins 
$1,...,N$. ${\cal V}_{\vec{\alpha}}$ maps the spins $1,...,N$ to the 
spins at the $N$ successive sites with orbitals up, where $\alpha_l=1$ 
and the spins $N+1,...,2N$ to the remaining $N$ successive sites with 
orbitals down, where $\alpha_l=-1$.
Using the cyclic invariance and this decomposition we obtain
\begin{eqnarray}
c_p&=&
{\cal C}
\left(
{\cal V}_{\vec{\beta}}^{\dagger}\:{\cal V}_{\vec{\alpha}}~{\cal P}^{(1)}
~R_{-p}^{(1)}R_{-p}^{(2)}~{\cal P}^{(1)}
\right)
\nonumber\\
&=&
{\cal C}
\left(
{\cal V}_{\vec{\beta}}^{\dagger}\:{\cal V}_{\vec{\alpha}}~{\cal P}^{(1)}
R_{-p}^{(1)}{\cal P}^{(1)}~R_{-p}^{(2)}
\right)
\nonumber\\
&=&
{\cal C}
\left(
{\cal V}_{\vec{\beta}}^{\dagger}\:{\cal V}_{\vec{\alpha}}
~R_{p}^{(1)}R_{-p}^{(2)}
\right).
\end{eqnarray}
The trace of the permutation ${\cal V}_{\vec{\beta}}^{\dagger}\:
{\cal V}_{\vec{\alpha}}R_{p}^{(1)}R_{-p}^{(2)}$ is graphically 
represented in Fig. \ref{fig:diagram1}A. The number of cycles $c_p$ 
is the number of closed loops in this diagram.

\begin{figure}[h!]
\begin{centering}
\vspace*{0.5cm}
\includegraphics[clip,width=1\columnwidth]{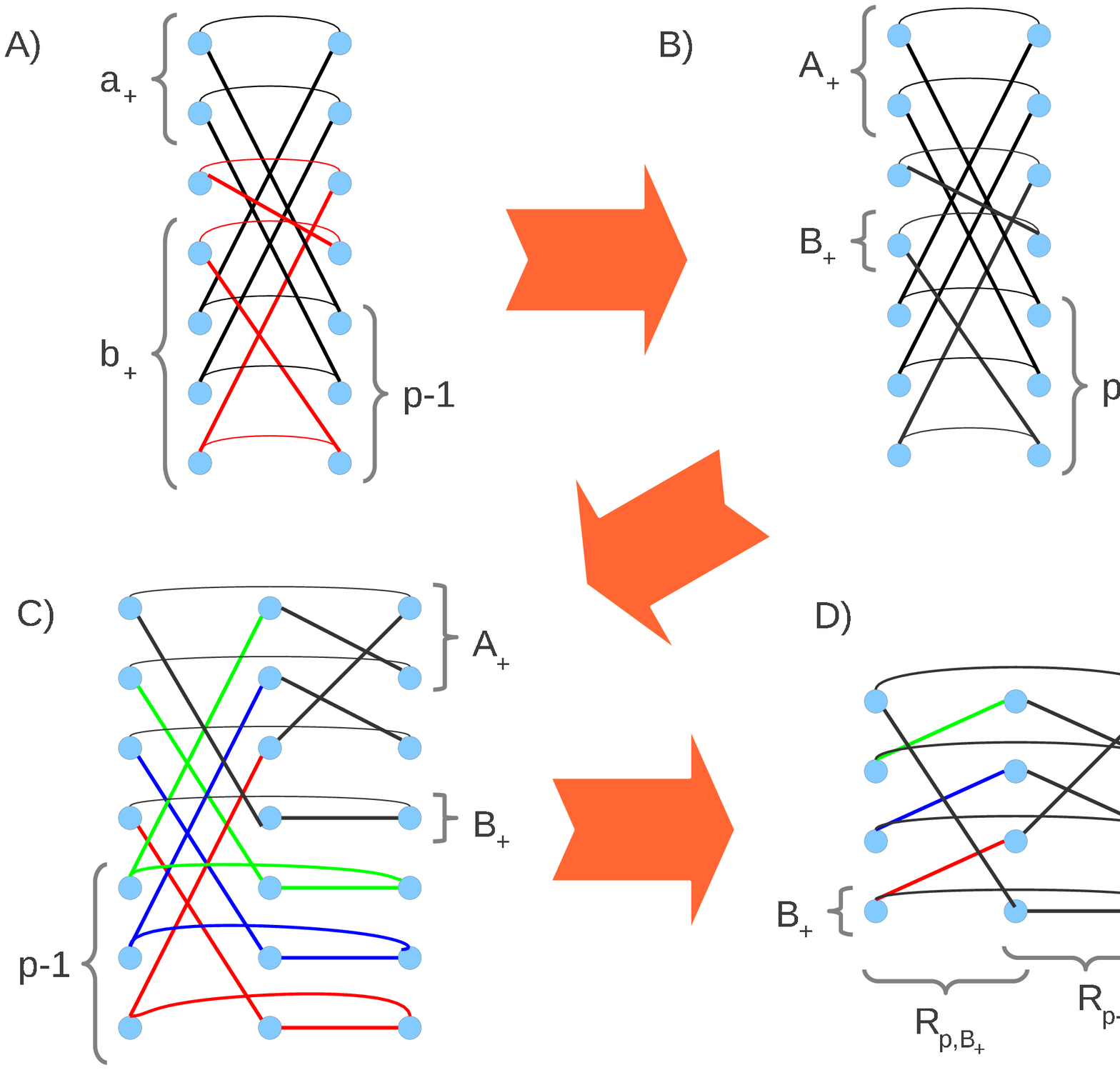} 
\par\end{centering}
\vspace*{-0.5cm}
\caption{ 
In A, 
the top half of the graph in Fig. \ref{fig:diagram1}B, but with the 
common loop connecting the top and bottom halves contracted to the top half. 
In B, 
the same as in A, but after application of the decompositions 
$a_+=C_+(p-1)+A_+$ and $b_+=(p-1)+D_+ p+B_+$ with integer $C_+,A_+,D_+,B_+$
and $|A_+|<p-1,|B_+|<p$. Here $C_+=D_+=0$ but in general, even when $C_+$ 
or $D_+$ are non-zero, the graph can be contracted to this form.
In C, 
graph B is transformed into a cyclic rotation of the first $A_++1$ spins 
by $A_+$ sites followed by a cyclic rotation of all 
$A_++B_++p$ spins by $-(p-1)$ elements. Notice the $p-1$ chains 
(marked green, blue, and red) that can be contracted to the $p-1$ single 
segments in graph D (marked green, blue, and red respectively).
In D, 
the final graph with the same number of closed loops as the top half of 
the graph in Fig. \ref{fig:diagram1}B including the common loop. Here a 
cyclic rotation of the first $p-1$ spins by $A_+$ sites is followed by a 
cyclic rotation of all $p$ spins by $B_+$ sites.
}
\label{fig:diagram2} 
\end{figure}

Figures \ref{fig:diagram1} and \ref{fig:diagram2} show a step-by-step 
reduction of the diagram \ref{fig:diagram1}A to a more and more compact 
form. The intermediate graph \ref{fig:diagram1}B can be split into the 
top and \hfill

%%%%%%%%%%%%%%%%%%%%%%%%%%%%%%%%%%%%%%%%%%%%%%%%%%%%%%%%%%%%%%%%%%%%%%%
%%%%%%%%%%%%%%%%%%%%%%%%%%%%%%%%%%%%%%%%%%%%%%%%%%%%%%%%%%%%%%%%%%%%%%%

\noindent
bottom halves [spins $1,...,N$ and $(N+1),...,2N$ respectively] 
that have one loop in common. In the following Fig. \ref{fig:diagram2}A, 
the common loop is included to the top half. Finally, the number of 
cycles (loops) in the top half, enlarged with the common loop, can be 
read from the graph \ref{fig:diagram2}D as
\begin{equation}
{\cal C}\left[R_{p,B_{+}}R_{p-1,A_{+}}\right].
\end{equation}
Here $R_{p,B_{+}}$ is a cyclic permutation of a $p$-element list by 
$B_{+}$ sites, and $R_{p-1,A_{+}}$ refers to the first $(p\!-\!1)$ 
elements of that list. In a similar way, the number of loops in the 
bottom half of the graph \ref{fig:diagram1}B, enlarged with the common 
loop, can be obtained as
\begin{equation}
{\cal C}\left[R_{p,B_{-}}R_{p-1,A_{-}}\right].
\end{equation}
When the two halves are put together again their number of cycles 
becomes a sum,
\begin{equation}
c_p=
{\cal C}\left[R_{p,B_{+}}R_{p-1,A_{+}}\right]+
{\cal C}\left[R_{p,B_{-}}R_{p-1,A_{-}}\right]-1,
\end{equation}
with the $-1$ to correct for the double counting of the loop common 
to both halves.

%\begin{thebibliography}{99}%%

%\bibitem{Kum13} B. Kumar, 
%                   Phys. Rev. B \textbf{87}, 195105 (2013).

%\end{thebibliography}

\eject

\end{document}